\documentclass[numsec,webpdf,modern,medium]{oup-authoring-template}
\onecolumn 
\graphicspath{{Fig/}}
\theoremstyle{thmstyleone}%
\theoremstyle{thmstyletwo}%
\theoremstyle{thmstylethree}%

\begin{document}

\journaltitle{Journal Title Here}
\DOI{DOI added during production}
\copyrightyear{YEAR}
\pubyear{YEAR}
\vol{XX}
\issue{x}
\access{Published: Date added during production}
\appnotes{Paper}

\firstpage{1}

\title[Spatial similarity in England]{Spatial similarity in socioeconomic data: a wavelet approach for England}
\author[1,$\ast$]{Duncan Cook}
\author[2]{John AD Aston}

\address[1]{\orgdiv{Statistical Laboratory}, \orgname{University of Cambridge}, \orgaddress{\street{Wilberforce Road}, \postcode{CB3 0WB}, \country{UK}}}
\address[2]{\orgdiv{Statistical Laboratory}, \orgname{University of Cambridge}, \orgaddress{\street{Wilberforce Road}, \postcode{CB3 0WB}, \country{UK}}}

\corresp[$\ast$]{Corresponding author. \href{mailto:email-id.com}{dc821@cam.ac.uk}}

\received{Date}{0}{Year}
\revised{Date}{0}{Year}
\accepted{Date}{0}{Year}

\abstract{Socioeconomic indicators in England exhibit complex spatial patterns that are not well captured by standard approaches based on averages or broad geographic classifications. We propose a method for comparing areas based on their internal spatial structure, using a multiresolution representation derived from the discrete wavelet transform.
The method embeds areal data into a regular grid, extracts local windows, and represents each as a set of scale- and direction-specific detail coefficients. A dissimilarity measure, defined over these coefficients and minimised over rotations and reflections, is used to identify contiguous sets of statistical units with similar spatial structure.
We apply the approach to England’s 2025 Index of Multiple Deprivation at the lower layer super output area level. We show that areas with similar internal structure are often found across regions, levels of urbanisation, and average deprivation, challenging the use of these categories as proxies for local geography.
The results provide a framework for identifying comparable places based on how deprivation is distributed within them, with implications for policy evaluation and transferability.} 


\maketitle

\section{Introduction}

Policymakers and researchers increasingly recognise that spatial inequality is not only about average differences between places, but about how variation is internally structured within them. The nuance of these spatial patterns was noted in recent flagship policy: the 2022 Levelling Up White Paper emphasised that for many indicators differences within regions can exceed those between regions \citep{levellingup}. Official releases of England's Index of Multiple Deprivation flag `pockets of deprivation' within every region, and warn that considering only higher levels of geography can hide variation in local conditions \citep{imd2025_research}. The implication for policymakers is to go beyond simple north-south or rural-urban stereotypes and to consider the small-area geographies of places where policy is delivered. 

A practical challenge is to do justice to local variation while still pragmatically deciding which places are genuinely comparable. Local nuance is important: hyper-local conditions contribute to the texture of a place. Two areas with the same average level of deprivation may present very different experiences for their residents depending on whether that deprivation is concentrated, dispersed, or spatially segregated. This internal structure matters for policy conclusions: in London, an area's economic inequality predicts stop-and-search activity even after adjusting for crime and demographics \citep{suss_stop_search}, and in US cities the association between income and violent crime depends on whether `local variation’ in income is defined as `within-neighbourhood', `between-neighbourhood', or `concentration' \citep{kang}. This suggests that an account of spatial structure has an impact on the transportability of a policy. 

\begin{figure}
    \centering
    \includegraphics[width=0.8\linewidth]{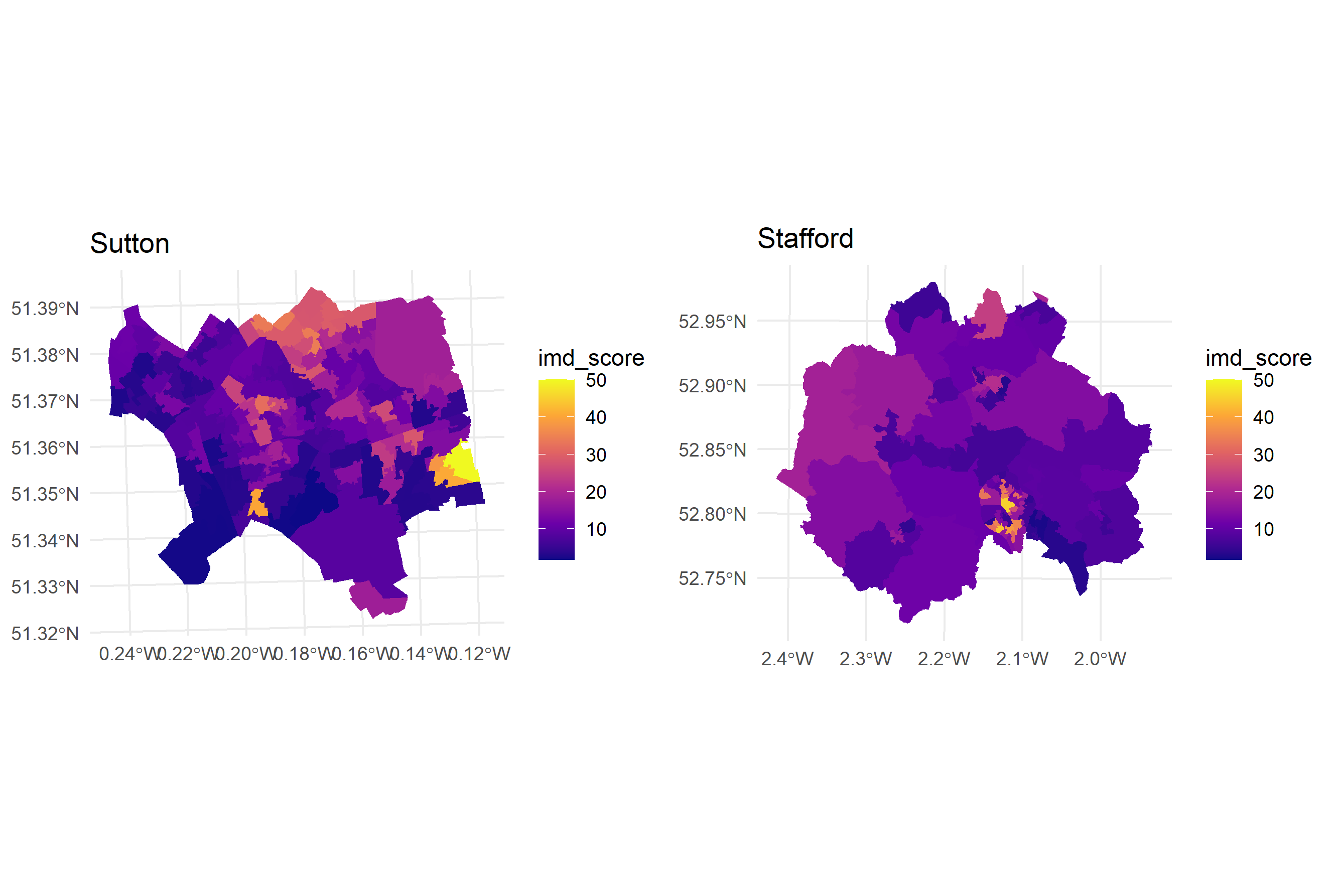}
    \caption{\textbf{Two local authorities with similar global spatial autocorrelation.} LSOA-level IMD scores for Sutton (left) and Stafford (right). Despite having very similar Moran’s I values (0.36) and similar mean deprivation, the spatial arrangements differ: Sutton has an east–west divide and several pockets of deprivation, while Stafford shows a smoother pattern with a small deprived centre.}
    \label{fig:same_morans_i}
\end{figure}

But there are good reasons to try to find ways in which places are alike: policy evaluation, benchmarking and peer learning all rely on some concept of `similarity'. Given the role of heuristics in decision-making \citep{tversky, heuristics}, analysts have a role in supporting policymakers by providing data-driven ways to group and compare areas that make the most of available data and methods. Recent system-wide user engagement at the 2025 UK Statistics Assembly \citep{stats_ass} affirmed demand for the production of granular statistics and tools to make their patterns visible, and in line with the Government Statistical Service's objective that data should be `subnational by default' \citep{gss}, low-geography data is increasingly available in the UK. Yet many policy tools handle spatial structure only indirectly. Definitions of statistical nearest neighbours that rely only on area averages, such as \citet{ons_cluster}, mask the underlying spatial structure within areas (see, for instance, Figure \ref{fig:same_morans_i}). Where analysis does attempt to incorporate within-area variation (e.g. using Moran’s $I$ to summarise local clustering, as in \citet{ons_moran} for income deprivation), this neglects where and at what scales variation concentrates, meaning commentary cannot reflect the geographical size of pockets of deprivation and whether they are in close proximity. For methods of policy evaluation where synthetic control areas are constructed as weighted composites of discontiguous areas \citep{wwscotland, xu}, although the trend in the average of an area's variable of interest can be closely approximated, the spatial structure is lost - and the comparator composite `area' does not have any meaning  (see \citet{britteon} for an example of public health devolution in England). This limits the usefulness of such analyses for peer learning. Public-facing services like ONS Explore Local Statistics \citep{ons_local} show a direction of travel, but there remains a gap for quantitative descriptions of internal structure.

This paper targets that gap by introducing a generalisable method for representing and comparing spatial structure in areal data. We make three contributions: 

\begin{enumerate}
    
    \item First, we propose that a wavelet representation of areal data can capture where variation concentrates and at which scales, and describe how this quantitative approach maps to qualitatively salient features of \textit{place} such as pockets of deprivation, differences between adjacent quarters, and north-south divides.
    
    \item Second, we enable comparison of the univariate spatial structure of places by defining a similarity index over pairs of this wavelet representation. By identifying the most important features of an area at multiple scales, we quantify the extent to which two areas share similar structures.
    
    \item Third, we apply this approach to England’s 2025 IMD, and reveal spatial similarities that defy simple heuristics for grouping local authorities (geographical region and rural-urban classification). As well as challenging the usefulness of these simple categories, we show that many local authorities do share spatial structure with other areas of England. 
    
\end{enumerate}

This approach provides a reproducible way to describe and compare the internal spatial configuration of an area: how variation is arranged in space and across scales, not just how large it is on average. It intends to support description and peer learning by challenging simple heuristics and providing a better alternative, giving a data-driven perspective that can be used when talking about univariate spatial variation in a `place'. While not itself a method for causal evaluation, it could be used as design triage: either upstream as an extra consideration for which areas to include in donor pools, or downstream to interpret where effects might plausibly propagate. This complements public-facing services which seek to make small-area data accessible for decision-making.


\section{Conceptual overview} \label{concepts}

We propose a method for representing and comparing small-area structure. We rasterise areal data to a regular grid (the British National Grid), producing a pixel map for England. The spatial distribution of a variable over a patch of land can then be represented by a matrix (here, standardised to dimension $2^K \times 2^K$ by aggregating cells to their mean in order to enable wavelet analysis). Where the area of interest is a collection of contiguous statistical units that together comprise an irregular shape (such as the LSOAs in a local authority), or the square patch of land includes missing data (such as in a lake or at a border), we impute missing values with the local mean at the smallest scale with sufficient coverage, iterating until all cells in the matrix are populated. We retain an explicit mask for the original area so that subsequent analysis uses only `true' data. 

For these standard matrices we then compute a wavelet representation to capture patterns of variation at multiple resolutions. We retain only the largest differences at each resolution by thresholding values to retain 90\% of the resolution-specific energy. The resulting representation is sparse and fast to compute while still retaining information about where variation occurs, and at what scale. Coarser resolutions capture large-scale differences (such as north–south contrasts within an area); finer resolutions capture `pockets' (adjacent neighbourhoods that differ sharply).

We are then able to quantify the spatial similarity between areas using this multi-resolution representation of spatial variation. We use the geometric mean of resolution-specific mean squared error of thresholded detail coefficients, averaging rescaled MSEs to ensure no resolution dominates because of the number of coefficients or due to the energy-preserving scaling in the transform. This places comparable weight on similarity at each resolution: pockets of variation are as important to capture as broad north-south gradients. For a given pair of square matrices we compute this score between the first matrix and the eight dihedral transforms of the second, and retain the most similar score. 

This lets us compare areas by the arrangement of variation within them, rather than only by averages or global summary statistics. Our analysis section shows how a user can define a pool of candidate areas (to avoid searching over the many square sub-areas of the large raster), and find the `most similar' candidate for a given focus area. Applied to the IMD we take local authorities as focus areas and demonstrate similarity across (not just within) the categories of region and urbanicity.


\subsection{Subnational data in England} \label{subnational}

We take advantage of the fact that aggregation of smaller statistical units yields an image-like representation of a geographical area, which can be analysed using statistical techniques that represent continuous surfaces as a sparse collection of key features. The analysis presented in this paper will use data available at the LSOA level, but the methodology is general enough to be used for data available at lower or at higher levels of aggregation\footnote{The smallest unit of statistical geography in England is an Output Area (OA). OAs are constructed by combining postcode building blocks within administrative areas, aiming to maximise social homogeneity within each OA and achieve similar population sizes across OAs (usually between 100 and 625 persons). Larger units of statistical geography then aggregate contiguous sets of smaller units. For example, a Lower Layer Super Output Area (LSOA) aggregates usually between four and six contiguous OAs, with a minimum population of 1000. Middle Layer Super Output Areas (MSOAs) aggregate LSOAs, and have a minimum population of 5000. \citet{martin} present the procedure for OAs in detail, and the ONS provides more background on census geographies \citep{ons_geog}.}. The chosen technique is introduced conceptually in the next subsection.


\subsection{Wavelets}

This paper proposes a wavelet approach to characterise geographic areas as a small number of important features at a range of scales. Wavelets succinctly capture the most important components of a function, distilling the information contained in a series or image into a sparse representation of only the most relevant features. The fundamental idea is that relatively few waves of different frequency, location and magnitude can represent even apparently complicated patterns. The process is used in diverse fields from extracting the signal in time series data to identifying the most important features in image processing.\footnote{The JPEG 2000 system for image compression, which converts information in the large number of pixels in an image into a small number of key features, is based on a discrete wavelet transform}

The name `wavelet', or `small wave', hints at a useful analogy: the surface of a body of water, while complicated, might be closely approximated by asking what combination of waves of different sizes or frequencies and at different locations together result in the surface: perhaps a few larger currents, some medium-sized waves, and a number of smaller ripples at particular locations. When applied to time series data, such as a series of economic activity, these different frequencies would correspond to longer wavelength patterns of boom and bust, seasonal patterns in different quarters, and differences in activity in different days of the week. 

We will use this technique to characterise the spatial distribution of a single variable over a geographic area. To fix ideas before formalising the approach, a wavelet decomposition of, say, a composite measure of deprivation would separate the overall pattern into changes in the average level at several scales, corresponding to currents, waves and ripples. Large, coarse changes in average deprivation between the north and south of a region would be captured by low resolution wavelets. The difference between two neighbouring large districts with different average levels of deprivation would be picked up by medium resolution wavelets. Changes between neighbourhoods, such as around pockets of deprivation in an otherwise affluent area, would be represented by larger values for fine-resolution wavelets in the relevant area. A region with a relatively constant north to south gradient of deprivation and little local variation might be represented by one relatively large low resolution wavelet and few or no medium or high resolution wavelets. A region with mostly uniform levels of deprivation, with one particularly affluent neighbourhood, might be represented by several fine-resolution wavelets at the relevant position, and few or no coarser-resolution wavelets.

As such, spatial variation around an average can be represented by a relatively small number of meaningful signals associated with a specific resolution, direction, location, and magnitude. This multiresolution approach allows the identification of `hyper-local' disparities without neglecting their broader context. Using this representation we will be able to identify, for a given area, similarly-shaped groups of contiguous statistical units that share the focus area's spatial variation at a number of resolutions.


\section{Methods} \label{methods} 


This section describes the procedure used to identify, for a given area of interest, a ranked list of contiguous statistical units with similar spatial structure. The method proceeds through a sequence of geographically interpretable steps.

First, areal data are rasterised onto a common grid, allowing polygon-level information to be treated as a continuous surface.

Second, square geographic windows of varying size are extracted. Within each window, values are aggregated, normalised relative to a local baseline, and missing cells are imputed to produce fully populated matrices suitable for comparison.

Third, each matrix is decomposed using a two-dimensional Haar transform \citep{mallat}, yielding multiresolution detail coefficients that capture spatial contrasts at different scales and orientations.

Fourth, coefficients below a threshold are removed, retaining only those features that contribute materially to spatial structure.

Finally, for each pair of areas, a scale-balanced dissimilarity metric is computed (minimised over rotations and reflections), producing a ranked list of candidate comparators. These are then mapped back to the underlying polygon geography, yielding contiguous sets of statistical units that correspond to interpretable places with their own internal spatial structure.


\subsection{Data preparation}

We rasterise LSOA-level values to a common grid in the British National Grid with cells 100\,m square. Each cell is assigned the value of the LSOA covering its centroid (i.e. no within-LSOA modelling). The 100\,m resolution balances fidelity and computational cost: LSOAs vary in geographical size, and the smallest are concentrated in densely populated urban centres. A 100\,m grid places at least one centroid in most LSOAs (capturing heterogeneity) while keeping transforms and distance calculations relatively fast.

More formally, let \(\{P_u\}_{u\in\mathcal{U}}\) be polygons (e.g. LSOAs in England\footnote{Polygon boundaries are published as shapefiles on data.gov.uk; see Section \ref{data} for the data used in this paper.}), and let \(Z_u\in\mathbb{R}\) denote the value (e.g. IMD score) associated with polygon \(P_u\).

Then consider the rectangular window \(W\subset\mathbb{R}^2\) which bounds \(\bigcup_{u\in\mathcal{U}} P_u\). Index cells in this window so that matrix \(C_{ij}\subset W\) is the 100m $\times$ 100m geographical footprint of a cell \((i,j)\) with centroid $c_{ij}$. Then the value of $C_{ij}$ is $Z_u$ for the $P_u$ containing $c_{ij}$ i.e., the cell takes the value of polygon that contains its centroid. Some cells may be missing if their centroids fall outside valid statistical polygons (e.g. bodies of water or neighbouring countries).


\subsection{Definition of a general area} \label{define_general_area}

We now wish to define a standard representation for a general patch of land in \(W\). The aim is to map any geographical area, regardless of its physical size or shape, onto a common spatial representation that preserves local patterns of variation while enabling like-for-like comparison across areas.

As such, we wish to construct a set of comparable square matrices of dimension $2^K \times 2^K$, representing patches of land in \(W\) (square submatrices of \(C\)) of a range of geographical sizes. For now, we describe a general \textbf{raw matrix} $A^{\text{raw}}$ of flexible size defining an area with reference to $C$, and describe how to generate its corresponding $2^K \times 2^K$ \textbf{analysis matrix}, $A$.

First allow $A^{\text{raw}}$ to be a square submatrix of $C$. Geographically, $A^{\text{raw}}$ corresponds to extracting a square `window’ of the underlying raster surface that fully contains the area of interest. It is defined by a `start cell' ($\tilde{i},\tilde{j}$) in $C$ to be the furthest north west cell of the represented area, and a geographic size for this area, $\omega \cdot 2^{K} \cdot$ 100\,m, which is the product of the width of the cell used for rasterisation (100\,m), $2^{K}$ where $K$ is the desired number of levels for the multiresolution analysis, and an integer aggregation factor $\omega$.\footnote{
Without loss of generality, this general area may be any set of contiguous statistical units, such as the set of LSOAs that make up a local authority. The raw matrix can be straightforwardly derived by identifying the smallest square of side $\omega \cdot 2^{K} \cdot$ 100\,m (with $\omega$ an integer) containing and centred on the bounding box of the region of interest.
}

As such, $A^{\text{raw}}(\tilde{i},\tilde{j},\omega,K)$ is a \(\omega \cdot 2^K \times \omega \cdot 2^K\) matrix where $A^{\text{raw}}_{ij} = C_{\tilde{i}+i-1,\tilde{j}+j-1}$ (which may be missing). Note that in analysis $K$ will fixed across all areas. We wish to obtain a fully populated \(2^K \times 2^K\) matrix for analysis from this raw matrix. To achieve this we apply the following steps of \textbf{aggregation} to control the size of the matrix, \textbf{normalisation} so that the matrix represents local variation, and \textbf{imputation} to populate missing values:

\begin{enumerate} 

    \item \textbf{Aggregation to a standard dimension matrix.} Define the \(2^K \times 2^K\) \textbf{aggregated matrix} $A^{\text{agg}}$ whose values are missing if more than half of the values in the corresponding \(\omega \times \omega\) matrix in \(A^{\text{raw}}\) are missing, and the mean of all nonmissing values if not. This standardises the size of the matrix object for analysis, while ensuring that each aggregated cell reflects a meaningful amount of underlying geographical coverage.

    \item To retain information about where data are genuinely observed, we also define the \(2^K \times 2^K\) binary \textbf{masking matrix} $M$ with $M_{ij}$ equal to 0 if $A^{\text{agg}}_{ij}$ is missing and 1 if not. The masking matrix will later be used to ensure that analysis comparing two matrices uses only their shared support. 
    
    \item \textbf{Normalisation for variation around a local average.} Calculate the mean of all non-missing values in $A^{\text{agg}}$ and call it the `local average', $c$. Define the \textbf{normalised matrix} $A^{\text{norm}}$ as the matrix that replaces each non-missing value in $A^{\text{agg}}$ with its value divided by $c$. This rescales the matrix so that its mean (over non-missing cells) equals ~1. Subsequent analysis can then focus on how values vary across space within an area.

    \item \textbf{Imputation with local averages for a fully populated matrix.} We then impute missing values, using the mean of local non-missing values from the smallest local submatrix that is more than half-populated. Iterating from small to large submatrices, this yields a fully populated matrix. First we define the unimputed $A^{(0)}$ as equal to $A^{\text{norm}}$, and then iterate over imputation levels $k\in [1,K-1]$ successively defining $A^{(k)}$ as:

    \begin{enumerate}
        
        \item If fewer than half of the cells in a $2^{k} \times 2^{k}$ submatrix of $A^{(k-1)}$ are empty, all empty cells are replaced with the mean of the values of all non-empty cells in that submatrix.
        
        \item If not, all cells in that submatrix remain unchanged.
        
    \end{enumerate}
        
    \item Then define the \textbf{analysis matrix} $A$ as the \(2^K \times 2^K\) matrix which replaces all empty cells in $A^{(K-1)}$ with the mean of all non-empty cells in $A^{(K-1)}$ (even if more than half of the cells are empty).

\end{enumerate}

$A$ is then the matrix of size $2^{K} \times 2^{K}$ made up of these imputed and aggregated cells. The aggregation step preserves information from the original $A^{\text{raw}}$ where there is sufficiently large geographical coverage; the normalisation step ensures variation is defined relative to the local average of pre-imputation cells (which is retained for later analysis), and the imputation step ensures a populated matrix while preserving averages at the finest resolution where there is sufficient information. 

The resulting analysis matrix $A$ therefore provides a fully populated, locally normalised representation of the area suitable for multiresolution decomposition. At its most general, this approach allows the identification of areas with similar spatial texture of a given variable, even when their overall average levels or geographic sizes differ.


\subsection{Wavelet transform} \label{wavelet}

For this analysis matrix representing a general square geographical area we now seek to define a representation that separates local variation from local average at different levels of resolution. The wavelet transform reorganises hierarchically the information in $A$, separating at each resolution $\ell$ the local average of blocks of adjacent cells as \textbf{approximation coefficients} and the variation within local blocks of cells into \textbf{detail coefficients} in three directions, $z$: left/right ($z=1$), up/down ($z=2$), and diagonal ($z=3$). 

At higher (finer) resolutions, the approximation coefficients are increasingly granular, and detail coefficients represent the difference between, say, adjacent neighbourhoods. These capture local variation that will not be picked up at coarser resolutions, but which will identify areas with dissimilarity between adjacent neighbourhoods. At lower (coarser) resolutions, approximations represent averages over large sections of an area, and details correspond to coarse variation between, say, the north and south.  

Formally, for our \(2^K \times 2^K\) analysis matrix \(A\), the two-dimensional discrete Haar transform following Mallat's pyramidal algorithm \citep{mallat} is performed with the following steps:

\begin{enumerate}

\item 
For ease of notation define $s^{(K)}=A$, the analysis matrix, such that $s^{(K)}_{i,j} = A_{i,j}$.

\item 
For each $\ell=K-1, \dots, 0$, recursively define the $2^\ell \times 2^\ell$ \textbf{approximation matrix} $s^{(\ell)}$ and three \textbf{detail matrices} $d^{(\ell),1}$, $d^{(\ell),2}$ and $d^{(\ell),3}$ where:

\begin{equation}
s^{(\ell)}_{ij} = 
\frac{1}{\sqrt{2}}
\Bigl(
    s^{(\ell+1)}_{2i-1, 2j-1} + s^{(\ell+1)}_{2i-1, 2j} +
    s^{(\ell+1)}_{2i  , 2j-1} + s^{(\ell+1)}_{2i  , 2j}
\Bigr)
\end{equation}

\begin{equation}
d^{(\ell),1}_{ij} = 
\frac{1}{\sqrt{2}}
\Bigl(
    (s^{(\ell+1)}_{2i-1, 2j}   +s^{(\ell+1)}_{2i,2j}) -
    (s^{(\ell+1)}_{2i-1, 2j-1} +s^{(\ell+1)}_{2i,2j-1})
\Bigr)
\end{equation}

\begin{equation} 
d^{(\ell),2}_{ij} =
\frac{1}{\sqrt{2}}
\Bigl(
    (s^{(\ell+1)}_{2i  , 2j-1} + s^{(\ell+1)}_{2i  , 2j}) -
    (s^{(\ell+1)}_{2i-1, 2j-1} + s^{(\ell+1)}_{2i-1, 2j})
\Bigr)
\end{equation}

\begin{equation} 
d^{(\ell),3}_{ij} =
\frac{1}{\sqrt{2}}
\Bigl(
    (s^{(\ell+1)}_{2i-1, 2j-1} + s^{(\ell+1)}_{2i  ,2j}) -
    (s^{(\ell+1)}_{2i  , 2j-1} + s^{(\ell+1)}_{2i-1,2j})
\Bigr)
\end{equation}

\item The $2^K \times 2^K$ matrix $H$ is then formed by assembling the approximation and detail coefficients into the standard 2D Haar layout. 

Define $H^{(0)} = s^{(0)}$, and for each $\ell = 0, \dots, K-1$ define:

\begin{equation} 
H^{(\ell+1)} = 
\begin{bmatrix}
    H^{(\ell)} & d^{(\ell),1}\\
    d^{(\ell),2} & d^{(\ell),3}
\end{bmatrix}
\end{equation}

We then set $H=H^{(K)}$.

\end{enumerate}

Applying this transform yields a multiresolution representation in which each coefficient is indexed by resolution $\ell$, spatial location $(i,j)$, and direction $z$, and encodes the magnitude and sign of local spatial contrast. Note that due to aggregation and normalisation, coefficient magnitudes vary systematically by resolution; we return to this in Section \ref{sec:distance}.


\subsection{Denoising} \label{denoising}

The distance described in Section \ref{sec:distance} is based on mean squared differences between detail coefficients at each resolution. To prevent numerous negligible residual coefficients from inflating the MSE, we apply a light denoising procedure prior to computing distances.

When there are missing values in either the focus area or a member of the universal pool, some detail coefficients will have been derived from imputed values. Because the imputation algorithm replaces missing values with local averages, such coefficients tend to be small; however, as they do not correspond to observed data, they are excluded via resolution-specific masking.

For each resolution $\ell$, we start by defining the $2^\ell \times 2^\ell$ masking matrix $M^{(\ell)}$ derived from the original mask $M$. Let $\bar{M}^{(\ell)}$ denote the matrix of block averages of $M$, where each cell is the mean of the corresponding non-overlapping $2^{K-\ell} \times 2^{K-\ell}$ block in $M$:

\begin{align}
\bar{M}^{(\ell)}_{uv}
&= \frac{1}{4^{K-\ell}}
  \sum_{i=2^{K-\ell}\cdot (u-1) + 1}^{2^{K-\ell}\cdot u}
  \sum_{j=2^{K-\ell}\cdot (v-1) + 1}^{2^{K-\ell}\cdot v}
  M_{ij}
\end{align}

We then define: 

\begin{align}
M^{(\ell)}_{uv}
&= \mathbf{1}\!\left\{
    \bar{M}^{(\ell)}_{uv} \ge \tau_\ell
  \right\}.
\label{eq:masking_matrix_ell}
\end{align}

where: 

\begin{equation}
\tau_\ell =
\begin{cases}
\frac{1}{2} & \text{if } \ell < K-2 \text{ (fine)},\\
\frac{1}{4} & \text{if } \ell \ge K-2 \text{ (coarse)}.
\end{cases}
\end{equation}

These thresholds ensure that, in general, detail coefficients are only used for subsequent analysis where they correspond to positions where at least half the cells in the analysis matrix were observed data. This threshold is relaxed at coarser resolutions: this prevents areas of irregular shape being left with very few coefficients.

For each direction $z \in \{1,2,3\}$, we define masked detail coefficients\footnote{For simplicity, masked coefficients are set to zero rather than treated as missing, and excluded from later analysis via the shared index set defined in Section \ref{sec:distance}.}:
\[
\tilde d^{(\ell),z}_{ij}
=
d^{(\ell),z}_{ij} \cdot M^{(\ell)}_{ij}.
\]

At each resolution $\ell$, we define $S^{(\ell)}$ as the smallest set of masked coefficients whose squared magnitudes account for $90\%$ of that level’s total masked energy:
\[
\sum_{(i,j,z)\in S^{(\ell)}}
\left( \tilde d^{(\ell),z}_{ij} \right)^2
\;\ge\;
0.9
\sum_{i,j,z}
\left( \tilde d^{(\ell),z}_{ij} \right)^2.
\]

and define \textbf{retained coefficients} as:
\[
r^{(\ell),z}_{ij}
=
\begin{cases}
\tilde d^{(\ell),z}_{ij} & \text{if } (i,j,z) \in S^{(\ell)},\\
0 & \text{otherwise.}
\end{cases}
\]


\subsection{Distance} \label{sec:distance}

We have now defined a general area $A$ (and its mask $M$ and local average $c$) with cells $A_{ij}$ representing $\omega$ x 100\,m squares on the British National Grid. Data from $C$ is used where there is sufficient coverage, and where this is not the case values are imputed in a way that preserves local averages. For each area we now have retained coefficient matrices 
$r^{(\ell),z}$ at each resolution $\ell$ and direction $z \in \{1,2,3\}$, as defined in Section~\ref{denoising}. 

We now construct a dissimilarity metric between two areas $p$ and $q$ based on these retained representations, so that for a given focus area and a user-defined universal pool of $N$ candidate comparator areas we can obtain a ranked list of scores for each of the areas in the universal pool. A pair of areas will be considered similar if, at each resolution, their retained coefficients are of similar magnitude across directions and positions. In particular, for two potential pairs where one pair matches well at all resolutions and the other matches better at all but one resolution and poorly in the other, we wish the former to be assessed as a `better' match.

\paragraph{Area-level coverage.}

Before computing distances, we exclude pairs of areas with insufficient spatial overlap. Let $M_{p,ij}$ and $M_{q,ij}$ denote the $2^K \times 2^K$ masking matrices for areas $p$ and $q$ defined in Section~\ref{define_general_area}. We define their shared coverage as the proportion of grid cells observed in both areas: 
\[
\kappa(p,q)
= \frac{1}{4^K}
  \sum_{i,j}
  M_{p,ij}
  M_{q,ij}.
\]
We exclude pairs with $\kappa(p,q)$ below a specified threshold, set in this analysis to $0.8$. This avoids computing distances for pairs that have only a small overlap of true data.

\paragraph{Resolution-specific shared support.}

At resolution $\ell$, define the shared coefficient index set
\[
\mathcal{I}^{(\ell)}_{p,q}
=
\left\{
(i,j,z) :
M^{(\ell)}_{p,ij}=1,
\; M^{(\ell)}_{q,ij}=1,
\; z\in\{1,2,3\}
\right\}.
\]
This indexes coefficients whose underlying spatial blocks meet the coverage threshold in both areas (for each direction).

\paragraph{Coefficient-level comparison}
Distance is computed using the retained detail coefficients  $r^{(\ell),z}_{ij}$ over the shared coefficient index set $\mathcal{I}^{(\ell)}_{p,q}$. We define for each resolution $\ell$:

\[
d_\ell(p,q)
=
\frac{1}{|\mathcal{I}^{(\ell)}_{p,q}|}
\sum_{(i,j,z)\in \mathcal{I}^{(\ell)}_{p,q}}
\left(
\frac{
r^{(\ell),z}_{p,ij}
-
r^{(\ell),z}_{q,ij}
}{
2^\ell
}
\right)^2.
\]

For a given scale–location–direction triplet in the shared support:

\begin{itemize}
    \item If both coefficients are zero, the pair contributes zero to the distance, reflecting that neither area has a feature present for that resolution-position-direction triplet.
    \item If one coefficient is zero and the other non-zero, the pair contributes a penalty, reflecting a feature present in one area but absent in the other.
    \item If both coefficients are non-zero with similar magnitude and sign, the contribution to distance is small.
    \item If both coefficients are non-zero but differ in sign, the contribution is large, reflecting structurally opposing spatial contrasts.
\end{itemize}

Similarity therefore reflects agreement in magnitude, direction, and location of spatial contrasts, as well as agreement in where contrasts are absent.

The division by $|\mathcal{I}^{(\ell)}_{p,q}|$, the number of resolution–location–direction coefficients  in the shared index set, expresses the sum of square differences in `per coefficient' terms\footnote{The restriction that $\kappa(p,q)$ exceeds a threshold ensures that $|\mathcal{I}^{(\ell)}_{p,q}|$ is nonempty.}. This ensures that fine resolutions do not dominate the overall score simply because they contain more terms. 

The division by $2^{\ell}$ adjusts for the scaling behaviour of the two-dimensional Haar transform: it offsets the inflation of amplitudes at coarse levels to ensure that MSEs across resolutions are of comparable magnitude. 

We then aggregate across levels with a geometric mean:
\[
D(p,q)
  = \left[
      \prod_{\ell = 0}^{K-1}
        \bigl(1 + d_\ell(p,q)\bigr)
    \right]^{1/K} - 1.
\]

A dissimilarity score of zero would correspond to identical retained coefficients at all resolutions.

Conceptually, the formulation treats mismatches in broad gradients and fine texture as equally salient components of spatial difference. The resolution-specific scaling places magnitudes on a comparable scale, and the geometric mean prevents strong agreement at one scale compensating for poor agreement at another. 

For a given pair, to remove dependence on map orientation, we minimise $D(p,q)$ over the dihedral-8 group $\mathcal D_8$ (rotations and reflections) and retain the minimum:

\[
D^\star(p,q)=\min_{g\in\mathcal D_8} D\bigl(g(p),q\bigr).
\]

By construction, the metric is invariant to overall average level (removed via normalisation in Section~\ref{define_general_area}) and to orientation (via minimisation over the dihedral group $\mathcal D_8$).


\subsection{Mapping back to geography}

Selected candidate comparators can then be identified in the raster as a square window together with a transformed binary mask derived from the focus area. To recover an interpretable polygon geography, we map this transformed mask back onto LSOAs intersecting with the matched window.

For a given candidate window, we first identify all LSOAs whose geometry intersects the window’s bounding box. We then align the transformed mask to the corresponding grid of aggregated raster cells and, for each intersecting LSOA, compute the proportion of relevant cells covered by the mask. This yields, for each LSOA, an overlap fraction between 0 and 1.

We then retain LSOAs whose overlap score exceeds a chosen inclusion threshold (in this case, $0.4$). The resulting set of polygons provides an interpretable approximation to the matched area in the original geography.


\subsection{Output} \label{output}

For a given \textbf{focus area} ($p$) and \textbf{universal pool} ($Q$) of $N$ areas ($q \in Q$), we then produce the \textbf{ranked list} for all $D^\star(p,q)$, with a lower score meaning closer similarity. Scores are mean-independent by default: we compare internal structure rather than levels, so two areas can be close on structure while differing in their averages. Optionally, the ranked list can be restricted to areas with similar means or geographical sizes.

This universal pool must be user-defined. As there is an infinite number of such areas on a finite continuous surface, we suggest some restrictions to define a computationally tractable search space of candidate areas; those used in the example analysis that follows are reported in Section \ref{sec:param-settings}.

We suggest a set of areas comprising sets of square areas (or `windows') forming a fishnet partition of the BNG-aligned 100m-resolution raster of England. The size of the partitions must be constrained to be a multiple of a $2^{K}$ (to enable wavelet decomposition) and one or more `aggregation factors' $\omega$. For example, an analysis that wanted to search over three-resolution representations of 300m and 500m square areas would correspond to partitions of England into 2400m and 3000m squares ($2^{3}$ times 300m and 500m respectively). This set could also include the smallest square areas containing any arbitrary regions of interest (such as local authorities).\footnote{This set can easily be extended depending on user needs. Further rotations, not aligned with $\mathcal D_8$ transforms of the British National Grid, can be introduced by rotating the raster. Different ranges of sizes can also be introduced by varying $\omega$. Users can also shift the origin of the fishnet over England (optionally many times per $\omega$), or rasterise to finer resolutions than 100m to include smaller areas as candidates in the universal pool.} This yields a set of matrices spanning the initial surface.

For a given area we are then able to look at whether its spatial variation is shared by other areas in England with different sizes or averages, and ask whether a given pattern is particular to, say, areas with the same rural-urban classification or in only that region. We are also able to map areas with few or no similar other areas, and ask whether they are in parts of England that may be considered in some sense unusual. 

The following section applies this analytical pipeline to local authorities in England.


\section{Parameter settings and reproducibility} \label{sec:param-settings}

\subsection{Focus areas: local authorities}

Local authorities are administrative regions of England governed by locally elected leaders, responsible for the provision of a range of services. Two common classifications used for grouping local authorities will be used for analysis: their geographical region, and their rural-urban classification. We will explore whether these two classifications are useful proxies when considering spatial variation. 

\begin{figure}
    \centering
    \includegraphics[width=1\linewidth]{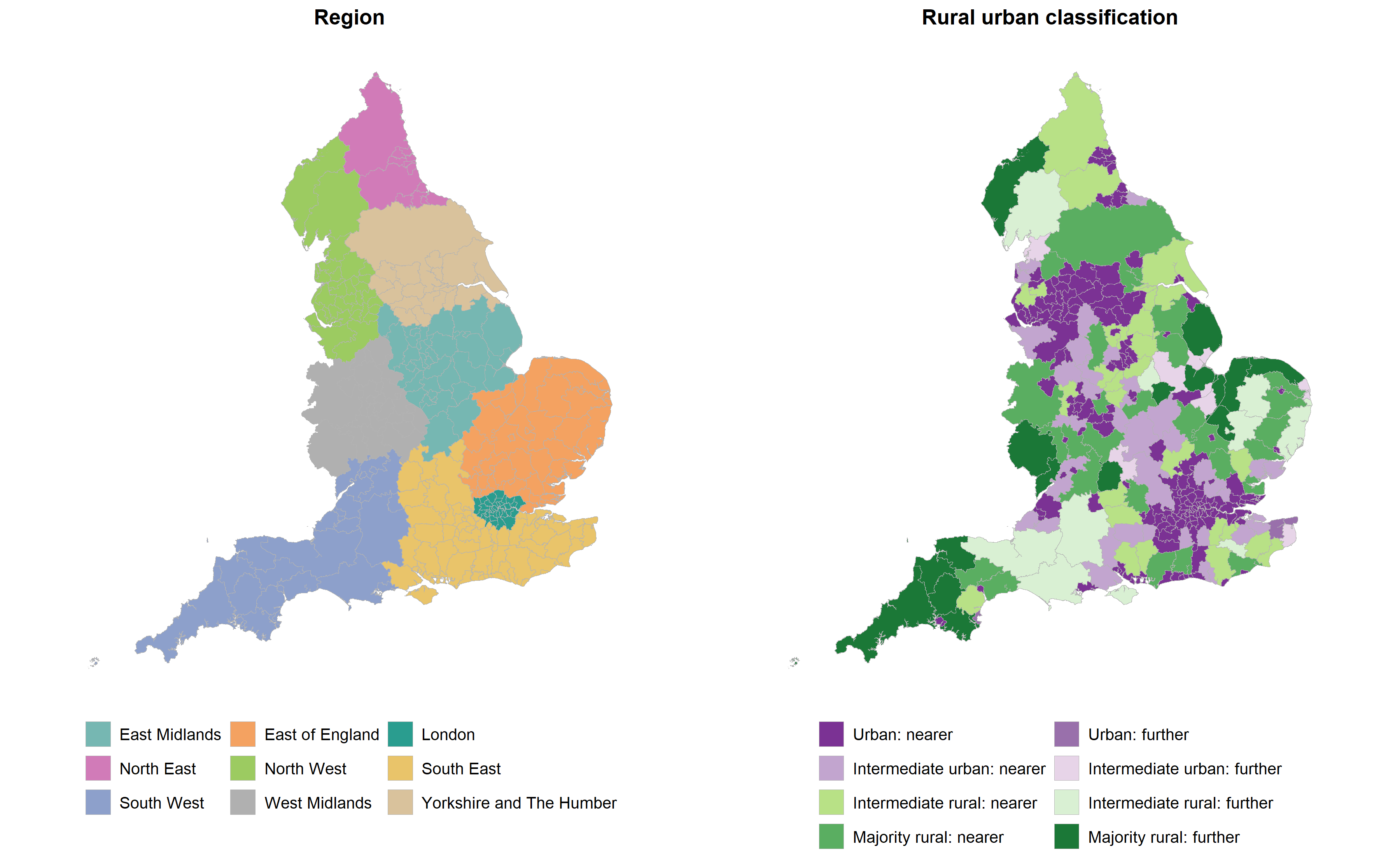}
    \caption{Classifications of local authorities in England. English local authorities are shown in the left panel by region and in the right panel by rural-urban classification.}
    \label{fig:locauth_maps}
\end{figure}

Figure \ref{fig:locauth_maps} shows local authorities grouped into regions, which are the highest tier of subnational division in England. The Office for National Statistics provides rural-urban classifications for local authorities that aggregate the rural-urban classification of their OAs, which are based on the size and density of the resident population of built-up areas. There are eight classifications, also mapped in Figure \ref{fig:locauth_maps}. We use these classifications as proxies for identifying similarity within and across recognised groupings of local authorities. Data availability is described in Section \ref{data}.

Given our choice of local authorities as the unit of analysis, we need to specify: (i) the wavelet depth $K$; (ii) inclusion thresholds for the number of LSOAs in each window of the universal pool; and (iii) the geographical sizes of windows for the universal pool, controlled by the aggregation factor.\footnote{The raw matrices for local authorities are straightforwardly derived by identifying the centre of their bounding box, and cropping to the smallest \(\omega \cdot 2^K\times \omega \cdot 2^K\) box in the BNG raster.} Our aim is to choose values that are interpretable, data-driven, and computationally tractable. In all cases we use simple rules of thumb combined with empirical distributions for English local authorities.

The steps below operate on a 100\,m raster of the British National Grid (EPSG:\allowbreak27700).


\subsection{Choosing the wavelet resolution} \label{sec:choose-K}

In selecting a value for $K$, we are trying to choose a maximum resolution large enough that it is able to capture variation between LSOAs, but not so large that at the finest resolution detail coefficients are mainly zero due to being in the interior of areal units. A maximum resolution that was too small would represent some LSOAs with many cells, and the detail coefficients in the interior of that LSOA would all be zero given values are constant within an areal unit. Conversely, a maximum resolution that was too large could have many LSOAs aggregated into one cell, blurring away the local variation we are trying to capture. 

To inform this choice we look at how a range of values of $K$ affects the distribution of the average number of cells per LSOA for local authorities in England. In practice our unit of analysis is the minimum bounding box for a local authority, and as local authorities are of irregular shape we calculate the `effective' number of LSOAs that box would contain if the area outside the local authority was tiled at the same LSOA density: $\tilde{N}_\alpha = \phi_\alpha N_\alpha$, where $N_\alpha$ is the number of LSOAs in local authority $\alpha$ and $\phi_\alpha$ is the reciprocal of the proportion that the area local authority $\alpha$ takes up of its bounding box. 

Then at resolution $K$ the average number of cells per effective LSOA is $C_{\alpha}(K) = \frac{4^{K}}{\tilde{N}_\alpha}$. If $C_{\alpha}(K)$ is much smaller than one, many LSOAs are aggregated into a single cell and between-LSOA variation is smoothed away. If $C_{\alpha}(K)$ is much larger than one, many cells represent one LSOA and the wavelet representation at fine resolutions will be dominated by zero coefficients corresponding to the interior of LSOAs. In practice we look for a $K$ such that $C_{\alpha}(K)$ is of order one for a typical LA, i.e. closer to $1$ than to $0.1$ or $10$.

\begin{figure}[tbp]
    \centering
    \includegraphics[width=0.5\linewidth]{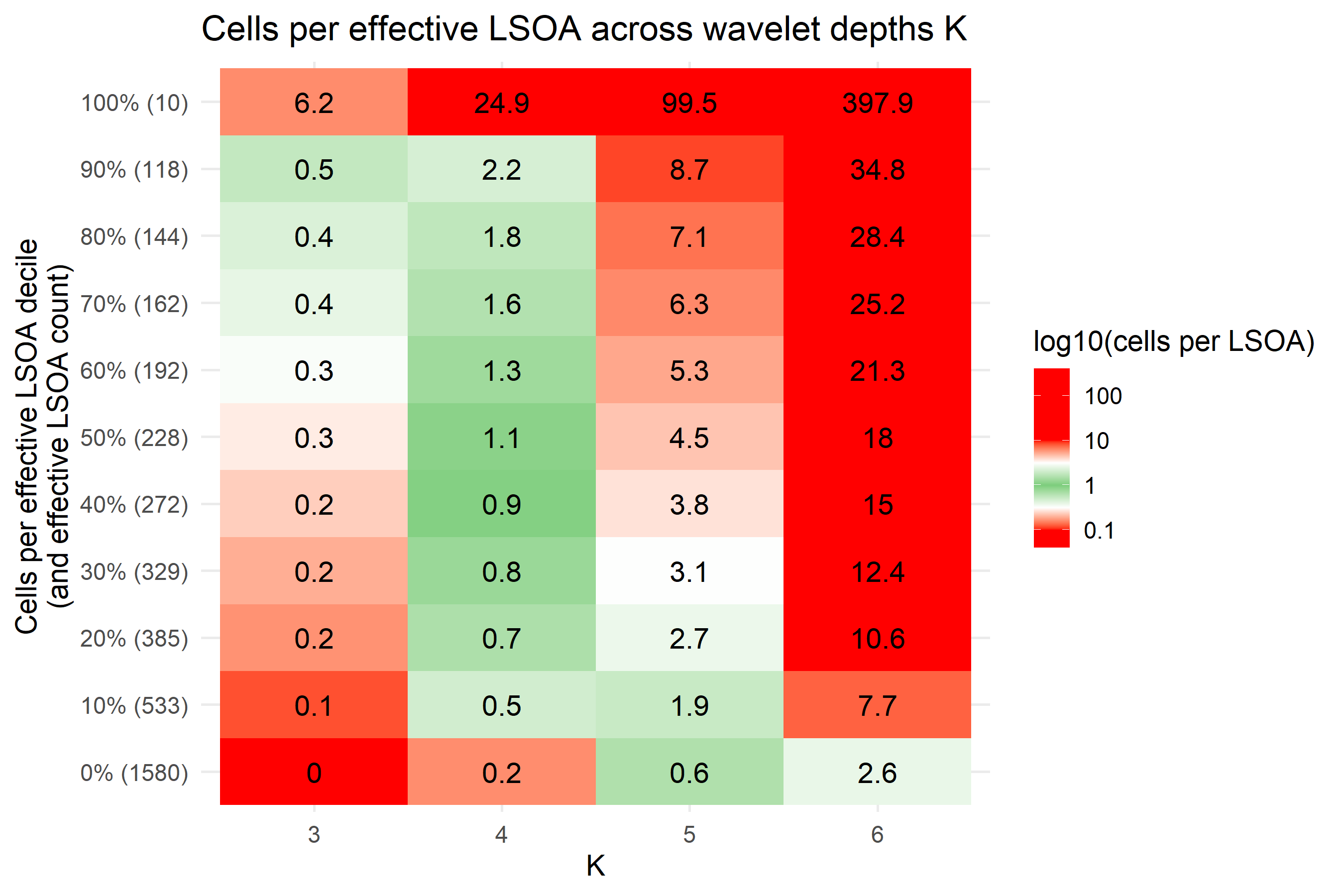}
    \caption{Average number of raster cells per effective LSOA at different K. Rows show deciles of the distribution across English local authorities. For small K many LSOAs are merged into single cells; for large K many cells fall within a single LSOA and fine-scale detail coefficients become dominated by zeros. K = 4 balances these extremes, with a central range of roughly 0.5–2 cells per LSOA.}
    \label{fig:cells-per-lsoa-vs-K}
\end{figure}

Empirically, the effective LSOA counts for English local authorities have a central 80\% range of $N^{\text{eff}} \in [118, 553]$ with a median of 228. Figure~\ref{fig:cells-per-lsoa-vs-K} shows the deciles of $C_{\alpha}(K)$ for $K=3,\dots,6$. For $K=3$ almost all deciles lie below one cell per LSOA, implying a coarse representation in which many LSOAs are merged into single cells. For $K=6$ the central deciles lie well above ten cells per LSOA, meaning many of the detail coefficients will be zero as an artefact of constant values within an LSOA.

By contrast, for $K=4$ the middle 80\% of authorities (10th-90th percentiles of $\tilde{N}$) have $C_{\alpha}(4) \in [0.5,2.2]$ cells per LSOA, providing a pragmatic compromise between over-blurring and over-specification. On this basis we fix the wavelet depth at $K=4$ for the remainder of the analysis.

\subsection{Choosing inclusion thresholds for the number of LSOAs per window}
\label{sec:choose-lsoa-thresholds}

Given $K=4$, we use the same logic to choose lower and upper bounds on the number of LSOAs per window in the universal pool. We want an LSOA to take up between roughly $0.5$ and $4$ cells, i.e. $C_{\alpha}(4) \in [0.5, 4]$ which for $K=4$ implies inclusion thresholds for the number of LSOAs in our window of $\tilde{N} \in [64, 512]$. 

The empirical distribution of $\tilde{N}_{\alpha}$ for English authorities is similar to this range, with the middle 60\% of local authority bounding boxes containing between 144 and 385 effective LSOAs. To exclude from our universal pool very small windows with very few LSOAs and very large windows with very many LSOAs, we adopt slightly more conservative thresholds and retain only windows containing between 100 and 400 LSOAs. 

\subsection{Choosing aggregation factors for the universal pool}
\label{sec:choose-aggs}

The final set of parameters controls the geographical sizes of windows in the universal pool. Each window is a square of side $2^{K} \times \omega \times 100\text{m}$, where $\omega$ is an aggregation factor determining how many original resolution cells are averaged into each cell before the wavelet transform. For $K=4$, a window therefore covers $1600 \times \omega$ metres in each direction.

To select a set of values for $\omega$ in our universal pool we take vigintiles of the implied aggregation factors for local authorities by dividing their maximum side length by $1600$ metres and rounding to the nearest integer. The resulting set of aggregation factors therefore spans the empirical range of local authority sizes. 

For each $\omega$ we tile England with a regular fishnet of windows of side length $2^K \times \omega \times 100$m, and apply a set of origin shifts to increase the sample size.

\section{Results} \label{results_imd}

In this section we apply the distance metric to the 2025 Index of Multiple Deprivation (IMD), an LSOA-level variable whose internal geography within a place is of policy interest. We show how the resulting scores should be interpreted, and demonstrate the kinds of claims that can be support.

We present the results in four stages: 

\begin{enumerate} 

\item First, we use a single local authority, Sutton, to show how the method translates LSOA-level data into a multiresolution representation and how distance scores correspond to visually interpretable similarities and differences in spatial pattern.

\item Second, still focusing on Sutton, we examine where its closest matches are found, comparing a pool calibrated to similar average deprivation with the full candidate pool.

\item Third, we extend the analysis to all local authorities in England, asking how common it is for authorities to have a close structural peer and whether relatively distinctive authorities show any obvious geographic pattern.

\item Finally, we test whether broad classifications, specifically region and rural-urban classification, are useful proxies for internal spatial structure by comparing, for each local authority, the best match within its own category with the best match outside it. 

\end{enumerate} 

Taken together, these stages move from interpretation of the method for one authority to a national assessment of the geography of structural similarity. They allow us to assess whether places that look similar on average deprivation, region, or urbanicity also share a similar internal geography of deprivation, and vice versa.

\subsection{Variable selection: the 2025 Index of Multiple Deprivation} \label{sec:imd_intro}

The Index of Multiple Deprivation is the official measure of relative deprivation for small areas in England. It is widely used in local planning, research, and national policymaking \citep{imd2025_research, imd2025_technical}, and continues to feature in decisions about funding to local government: both for identifying disadvantaged neighbourhoods to receive funding under the Pride in Place Programme \citep{pride_in_place}, and at the local authority level in proposed reforms to funding formulae \citep{ffr}. A new iteration was published in 2025, providing the first update since 2019. It combines indicators across seven domains of deprivation (income, employment, education, health, crime, barriers to housing and services, and living environment) into a single continuous score for each LSOA. Higher scores correspond to higher levels of deprivation. 

The IMD is a useful variable for demonstrating the proposed method because of its availability at small geographies and the policy interest in its spatial texture. Official documentation and related policy discussions emphasise that deprivation is not simply a matter of average levels across areas: in some places it is concentrated in pockets, while in others it is more evenly distributed \citep{imd2025_research}. 

We therefore treat the IMD as a representative spatial variable whose local and national patterns are of substantive interest, making it well-suited to demonstrating how the proposed approach characterises spatial texture in a policy-relevant variable. For the purposes of exposition, we use the continuous IMD score rather than ranks. To provide summary characteristics for candidate areas that do not correspond to administrative units, we assign region and rural-urban classifications based on the LSOA containing the centroid of each area. We use the `local average', $c$, (defined in Section \ref{define_general_area}) as a proxy for average deprivation level, and define two areas as having `similar average deprivation' if the absolute difference in the percentile positions of their values of $c$ is at most 0.1 (10 percentile points), using the empirical distribution of non-missing values in the rasterised IMD surface as the reference distribution.

Applying the method yields, for each focus area, a ranked distribution of dissimilarity scores over the universal pool of candidate areas. These scores form the basis of the analyses that follow. In all results, reported scores are minimised over rotations and reflections; where figures display untransformed areas, this is for visual interpretability only.


\subsection{Interpreting spatial similarity for a single authority} \label{results_imd_cases}

We begin with Sutton to illustrate how the method should be interpreted for a single authority before moving to national patterns. Sutton is used as an expository example for two reasons. First, its internal geography exhibits variation at multiple spatial scales, making it a useful case for demonstrating how the multiresolution representation captures both local pockets and broader gradients. Second, it follows directly from Figure \ref{fig:same_morans_i}, where Sutton was contrasted with Stafford as an example of two authorities with similar global spatial autocorrelation (Moran’s I) but different internal spatial structure. Using Sutton here for exposition allows us to connect the limitations of global summary measures to the richer representation developed in Section \ref{methods}.

The aim of this section is not to claim that Sutton is representative, but to provide a concrete example in which the objects defined in Section \ref{methods} can be linked to visually recognisable spatial features.

\begin{figure}
    \centering
    \includegraphics[width=0.9\linewidth]{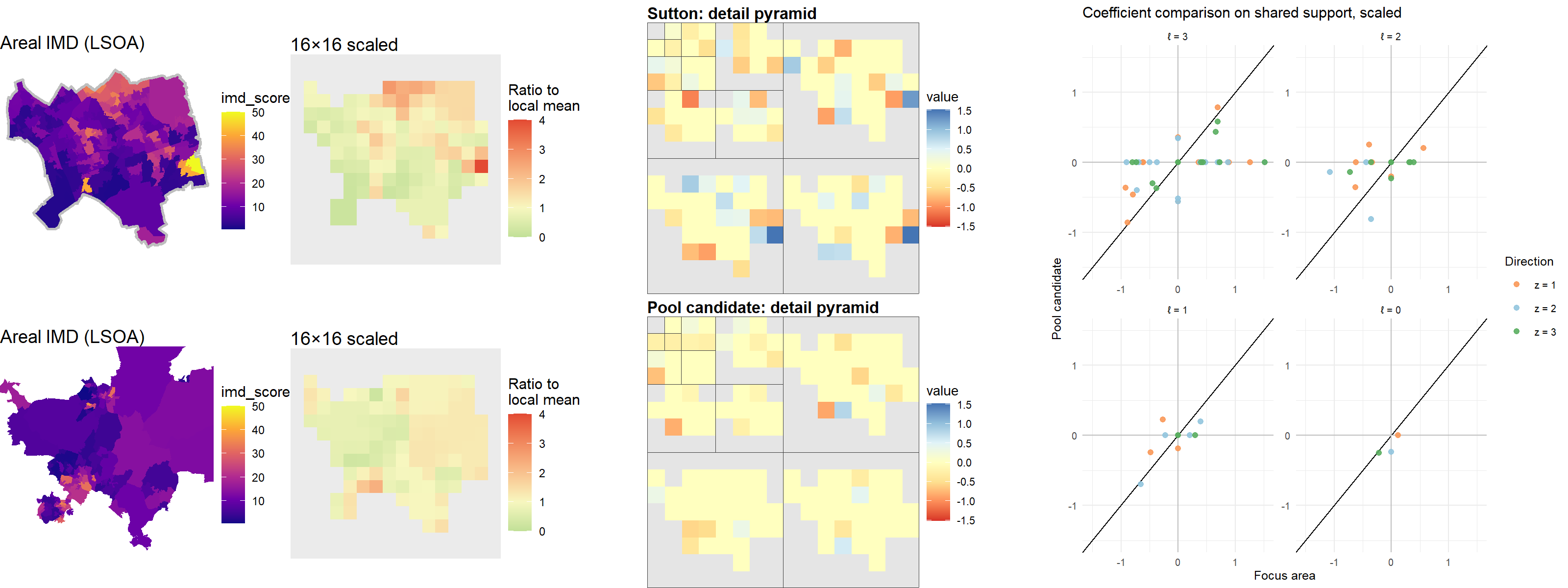}
    \caption{Sutton and its best match from the universal pool. The panels show, from left to right, the original LSOA geography for Sutton and the LSOAs retrieved by mapping back to administrative boundaries from the matched masked area, the $16 \times 16$ local variation matrices for each area, the thresholded detail coefficients by resolution and direction, and coefficient-by-coefficient comparisons. Detail coefficients are shown after the same resolution scaling used in the coefficient comparison and distance metric. In the scatterplot, general agreement can be seen across resolutions $\ell=0,\dots,K-1$. Several retained coefficients lie close to the $x=y$ line, indicating features of similar magnitude in the same location. Only a small number of retained coefficients lie in opposite-sign quadrants, indicating that the two areas rarely exhibit structurally opposing variation at the same resolution-location-direction triplet.}
    \label{fig:sutton_best}
\end{figure}

\begin{figure}
    \centering
    \includegraphics[width=0.9\linewidth]{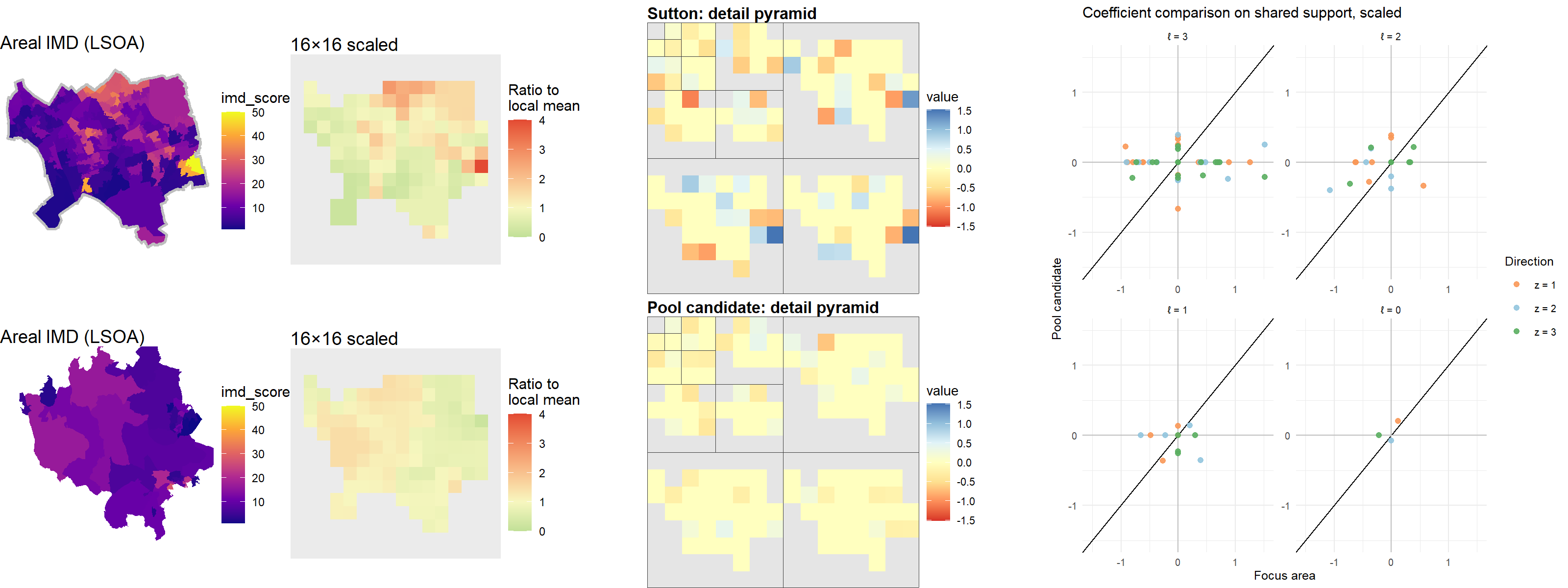}
    \caption{Sutton and a near median match from the universal pool. Panels are as in Figure \ref{fig:sutton_best}. In the scatterplot, unlike in Figure \ref{fig:sutton_best}, there is little agreement across resolutions. Few retained coefficients lie close to the $x=y$ line, and many points are found on the axes. This shows that the two areas rarely exhibit similar variation at the same resolution-location-direction triplet.}
    \label{fig:sutton_median}
\end{figure}

\begin{figure}
    \centering
    \includegraphics[width=0.9\linewidth]{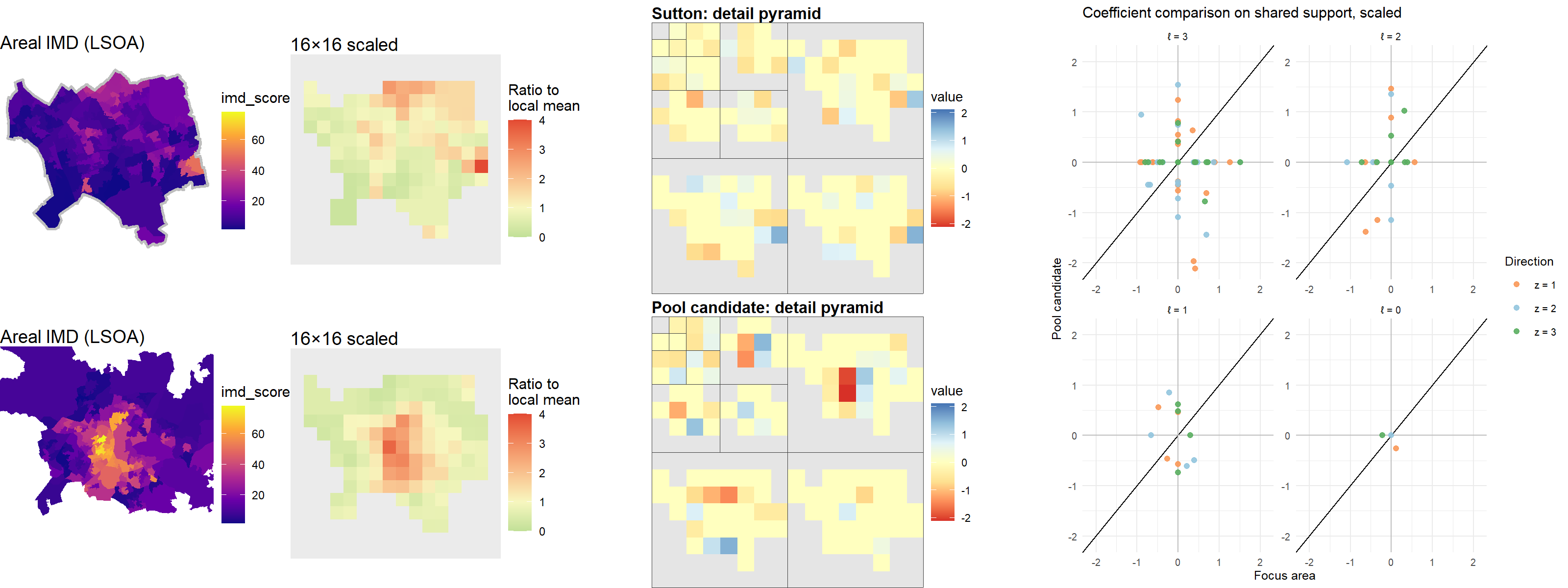}
    \caption{Sutton and a bad match from the universal pool. There is little correspondence between coefficients, and the areal data clearly show very different spatial structures.}
    \label{fig:sutton_worst}
\end{figure}

Figure \ref{fig:sutton_best} presents Sutton and its closest match from a mean-restricted pool: an area centred on Winchester with a distance score of 0.09. Matches are identified by minimising dissimilarity over all rotations and reflections (the dihedral group $D_8$), ensuring that orientation does not affect similarity. For clarity of visual comparison, however, the examples shown in Figures \ref{fig:sutton_best}, \ref{fig:sutton_median} and \ref{fig:sutton_worst} are restricted to matches in their original orientation, so that spatial patterns can be directly compared visually.

The left-hand panels show the original LSOA geography, shaded by IMD score. The second panels show the corresponding shape-masked $16 \times 16$ local variation matrices obtained after rasterisation, aggregation, and normalisation. The third panels show the thresholded wavelet detail coefficients by resolution and direction. The right-hand scatterplot compares the retained detail coefficients for the two areas, faceted by resolution.

Together these panels show how visually recognisable features are translated into the multiresolution representation. Sutton and its best match share several features across resolutions. At the finest resolution, the closest match picks up two of Sutton's three pockets of deprivation: those toward the south west, and toward the north. Broad patterns at middle resolutions share the general gradient of a more deprived south west, and at the coarsest resolution both display only small variation across quarters. In the right panel of Figure \ref{fig:sutton_best}, we see that across several resolutions, retained coefficients align in both sign and magnitude at the same spatial positions: these are points around the $x=y$ line in the scatterplot. Points at the origin correspond to positions where neither area has a non-thresholded feature. There are relatively few examples of the two areas exhibiting structurally opposing contrasts at the same scale and location (points in the top-left or bottom-right quadrants).

For comparison, Figures \ref{fig:sutton_median} and \ref{fig:sutton_worst} show the areas with distance scores around the median and at the poorest-matching end of the distribution of scores from a pool restricted to areas with a similar mean (in Warwick and Derby with scores of 0.13 and 0.36 respectively). The near-median match in Figure \ref{fig:sutton_median} exhibits relatively little retained variation; toward the poorer-matching end of the distribution, in Figure \ref{fig:sutton_worst}, the spatial configuration is clearly very different to Sutton's, despite varying around a similar local average. 

As such, a close match is not simply an area with a similar average deprivation level. It is an area whose internal geography generates similar retained variation across multiple scales.

\subsection{Spatial distribution of close matches}

Having established how a close match should be interpreted, we now consider where Sutton's close matches are found. The aim is to demonstrate for a concrete example that structurally similar areas need not be geographically proximate or confined to the most obvious comparator groups - nor necessarily of the same size or of a similar average level. 

\begin{figure}
    \centering
    \includegraphics[width=0.9\linewidth]{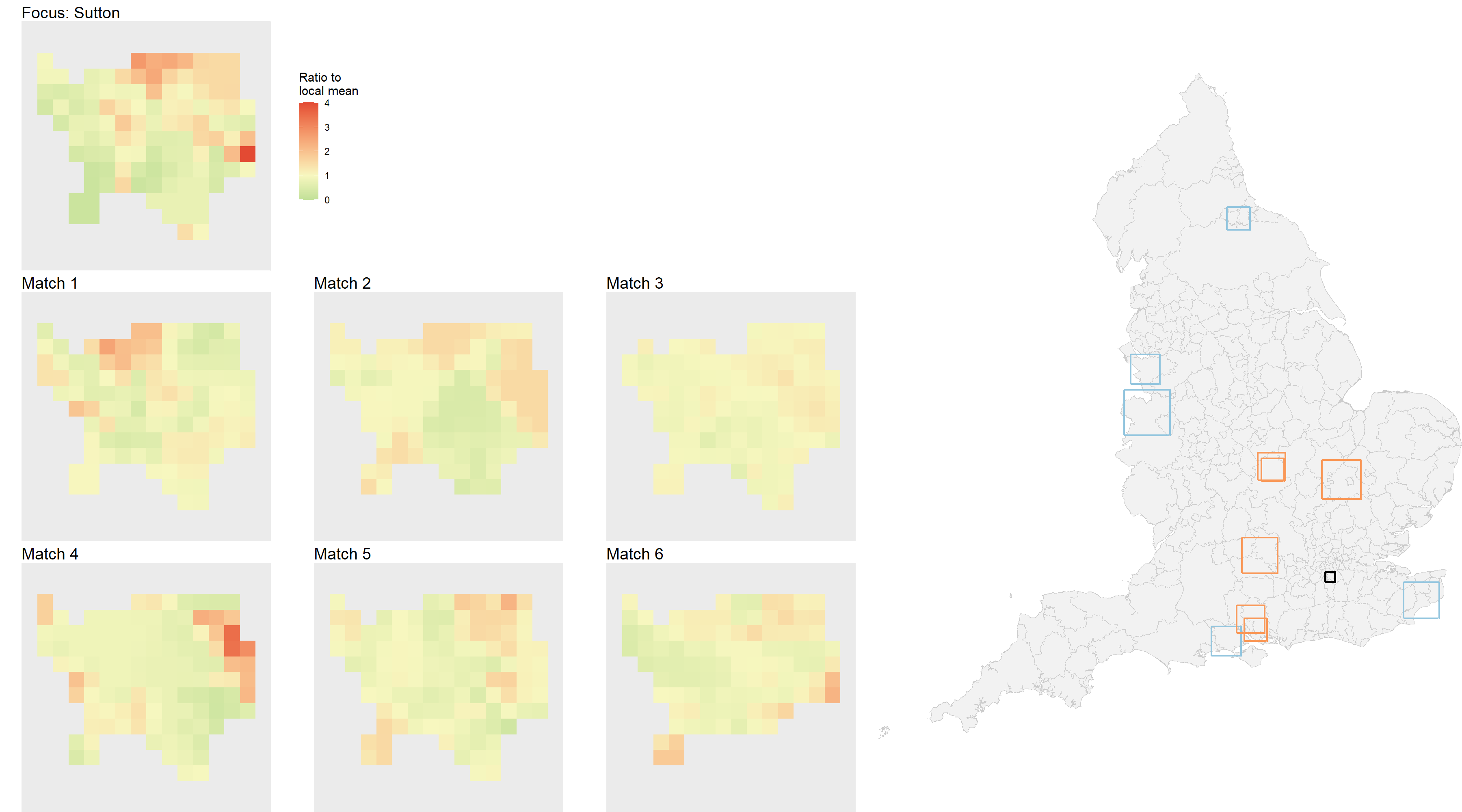}
    \caption{Sutton’s closest matches. The left panels show Sutton’s local variation matrix and the eight lowest-scoring matches from the unrestricted universal pool. The right panel maps the locations of the corresponding candidate windows across England: within a pool of candidates with average deprivation in the same decile (orange), and from an unrestricted pool (blue). Close matches are not geographically clustered around Sutton and are not confined to areas of similar size. This is true for matches with similar and unrestricted average levels.}
    \label{fig:Sutton_match_gallery}
\end{figure}

Figure \ref{fig:Sutton_match_gallery} shows Sutton's local variation matrix alongside those of its closest matches from the universal pool, and maps the locations of these matching areas across England. The map shows that close matches can be found across different parts of England: for Sutton, the closest matches are found in several parts of the country rather than concentrated in one obvious comparator group. The windows in orange correspond to the best matches from a pool restricted to areas in the same decile of average deprivation as Sutton, and the windows in blue are the best matches regardless of average level.

This illustrates, for a single concrete case, the broader claim that internal spatial structure is not well captured by simple heuristics such as region, urbanicity, size, or average deprivation level. A place may have structural peers in any region of the country, or that sit at a different average level of deprivation. The next subsection shows that this is not particular to Sutton, but is characteristic of the national pattern.

\subsection{Matchability across local authorities} \label{sec:matchability}

We now switch perspective from considering one specific local authority to considering all local authorities across England. 

A common approach is to understand `similarity' as meaning similar average levels of deprivation. We therefore begin with a calibrated search, restricting the candidate pool to areas whose local average deprivation lies in the same decile of the national distribution as the focus authority. This allows us to ask: even among areas with similar average deprivation, is it usually possible to find another area with a similar internal geography of deprivation?

\begin{figure}
    \centering
    \includegraphics[width=0.9\linewidth]{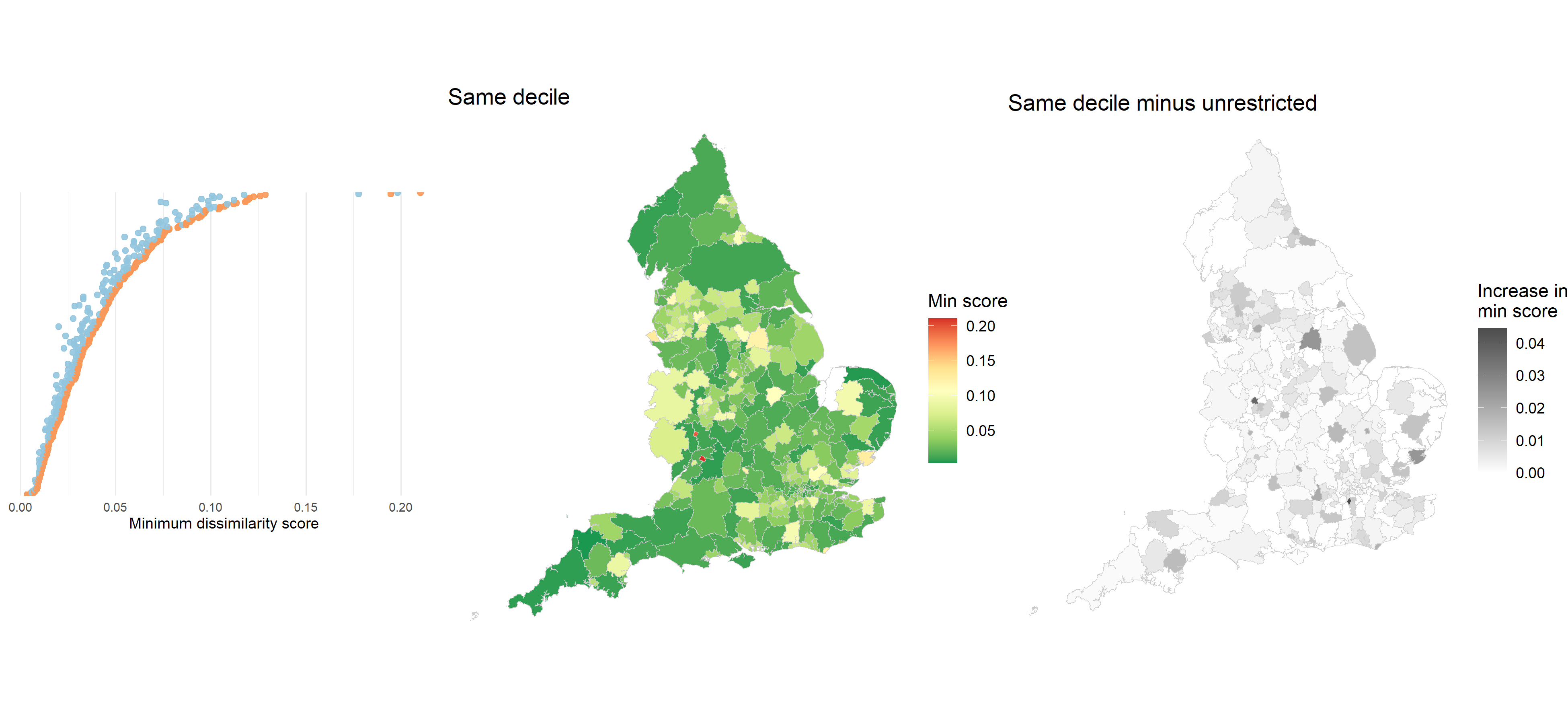}
    \caption{Minimum dissimilarity scores for local authorities in England. Left: minimum dissimilarity score for each authority under a same-decile calibrated search (orange) and an unrestricted search (blue), ordered by the calibrated minimum. Middle: map of calibrated minimum scores. Right: gain from removing the same-decile restriction, defined as the calibrated minimum minus the unrestricted minimum. Most authorities have a reasonably close structural peer even under calibration to similar average deprivation, and the gains from relaxing this restriction are usually modest and show no strong geographic pattern.}
    \label{fig:min_scores_dist_and_map}
\end{figure}

Figure \ref{fig:min_scores_dist_and_map} summarises matchability across England under both calibrated and unrestricted searches. The left panel shows, for each local authority, the minimum dissimilarity score obtained when the search is restricted to same-decile candidates, alongside the unrestricted minimum for comparison. Authorities are ordered by their restricted minimum\footnote{Because local authorities are typically irregular shapes, they vary in the degree to which they fully populate their bounding box. As such the number of valid coefficients contributing to the distance score varies across areas, particularly at finer resolutions. Authorities that occupy a smaller proportion of the analysis matrix therefore tend to have slightly lower minimum dissimilarity scores as their smaller number of coefficients makes their structure easier to match. Because potential matches are masked to the local authority's shape, the ranking of potential matches for a given authority is not affected, but it does imply that minimum scores are not strictly comparable across authorities with very different coverage.}. The middle panel maps the restricted minimum scores. 

Most authorities still have a reasonably close structural peer even under this restriction, and there is no strong visual evidence that authorities with relatively higher or lower mean-restricted minimum scores are concentrated in any one part of the country: they occur in the north and south, in urban and rural areas, and on the coast as well as inland.

Some authorities have higher calibrated minimum dissimilarity scores than others. Part of this variation reflects differences in how fully authorities occupy the analysis grid, especially at finer resolutions, so absolute minima should not be interpreted as pure measures of structural distinctiveness. Nonetheless, most authorities have reasonably close matches. The two highest calibrated minima are observed for Cheltenham and Worcester.

The right-hand panel then shows the gain from removing the same-decile restriction. By construction this gain is non-negative: enlarging the candidate pool cannot worsen the best available match. For many authorities the gain is modest, indicating that a close structural peer can often be found even within the stricter comparison set. This is clear from the distribution of gains: the median improvement from removing the restriction is only 0.002, and even the upper quartile is below 0.006, although a small number of authorities benefit more substantially. The largest gains are observed for authorities such as Epsom and Ewell and Wolverhampton. Again, these gains are not obviously geographically clustered.

From a policy perspective, this matters because restricting attention to same-average peers may exclude areas with closely comparable internal structure - and even if it is desirable to consider peer areas with similar average levels, it is often possible to find within that group areas that also share a similar internal structure. This suggests that when peer learning or policy transportability depends not only on how deprived a place is on average, but on how deprivation is arranged within it, it is possible to find comparator areas that meet a stricter definition of internal structure, and that these may be found across average levels of deprivation.

\subsection{Structural similarity and classification} \label{in_out_plots}

Finally, we test whether broad classifications are useful proxies for internal spatial structure. If region or rural-urban classification captured the geography of deprivation, then the closest structural match for a local authority would typically be found within its own category - for example, one might expect for an authority in the West Midlands that its best matching area would also be in the West Midlands. In that case, the best within-category match (`best-in') would outperform the best out-of-category match (`best-out').

For each local authority, we therefore compare the minimum dissimilarity score obtained within its own category with the minimum obtained outside it, considering both region and rural-urban classification. We summarise this comparison by:
\[
\Delta = -\bigl(D_{\text{in}} - D_{\text{out}}\bigr),
\]
so that positive values indicate that the best within-category match is closer than the best out-of-category match.

\begin{figure}
    \centering
    \includegraphics[width=1\linewidth]{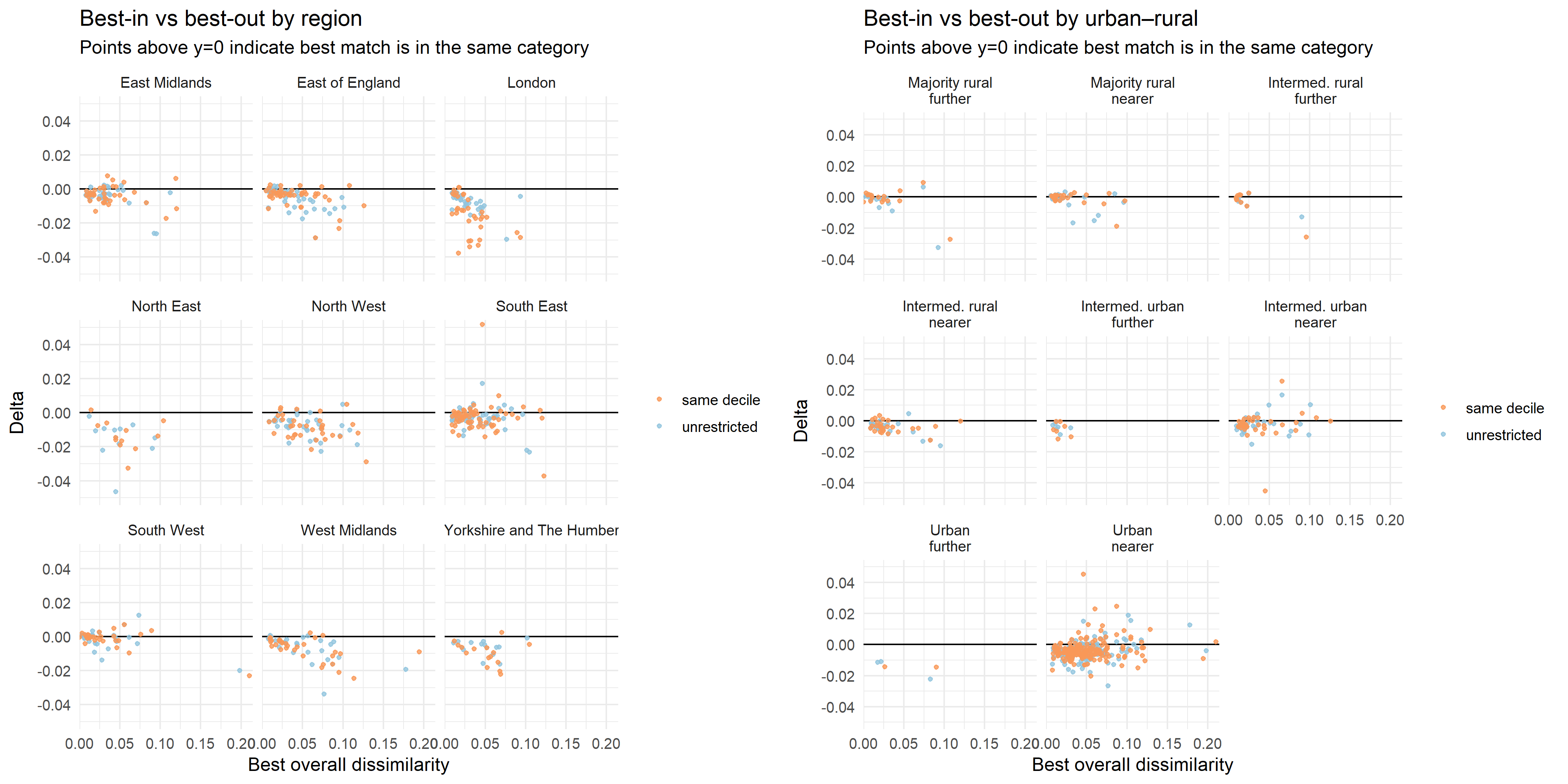}
    \caption{Best-in versus best-out comparisons by category. Each point is a local authority. Positive values of $\Delta$ indicate that the best within-category match is closer than the best out-of-category match. Blue points use the unrestricted pool; orange points use a pool calibrated to similar average deprivation. Across both region and rural-urban classification, values are concentrated at or below zero, indicating that the closest structural matches are often found outside the authority’s own category.}
    \label{fig:bestinout}
\end{figure}

Figure \ref{fig:bestinout} plots $\Delta$ against each authority’s best overall dissimilarity score, under both an unrestricted search and a search calibrated to similar average deprivation, for both region and for rural-urban classification. If these classifications were informative proxies for spatial structure, most points would lie above zero. Instead, across both region and rural-urban classification, the distribution of $\Delta$ is centred at or below zero - namely, the best structural match for a local authority is often found outside its own category. This remains true even after calibrating to similar average deprivation, indicating that the result is not driven by differences in average level.

As such, broad classifications are weak proxies for the internal spatial variation of an area. In the majority of cases, the best out-of-category match is better than any within-category match. This has two implications for policymakers thinking about transportability, benchmarking, evaluation, or peer learning. Firstly, assuming that two areas in the same broad classification are `similar’ is likely to neglect differences in their internal spatial variation. Second, and more positively, our results show that if policymakers do wish to identify similarity in patterns of spatial variation, our method provides an approach - and that these similar areas may be outside of usual classifications.


\section{Conclusion} \label{conclusion}

This paper has introduced a method for comparing the internal spatial structure of areal data. By rasterising irregular areas to a common grid, transforming them into a multiresolution wavelet representation, and comparing retained coefficients across scale, direction, and location, the approach provides a reproducible way to identify areas with similar spatial texture. The method is designed to capture how deprivation is arranged within an area at multiple resolution, allowing matching on variation at coarse levels through to smaller local pockets of deprivation.

Applied to England’s 2025 Index of Multiple Deprivation, the results show that in general, areas that have similar average deprivation level or share a region or rural-urban classification can have very different internal structures, challenging the use of simple classifications as proxies for spatial similarity. 

However, similar spatial structures can be found across regions, rural-urban classifications, and average levels of deprivation. In contrast to approaches that rely on weighted composites of discontiguous areas, our method identifies contiguous areas with similar internal structure, meaning that the final matched area remains geographically interpretable as a place. As such it provides a complementary approach to thinking about similarity that captures the geography of local variation. More broadly, we argue that our method can help analysts and policymakers draw a link between granular areal data and the within-area spatial variation that matters for place-based policymaking.


\section{Data availability}\label{data}

The 2021 Lower Layer Super Output Area (LSOA) population-weighted centroids for England and Wales, as published in December 2025, are available from \href{https://www.data.gov.uk/dataset/e3e903a6-1864-4083-8837-017b6bdf8cc5/lower-layer-super-output-areas-december-2021-ew-population-weighted-centroids2}{data.gov.uk}.

The 2021 Rural Urban Classification (RUC21) is available from the Office for National Statistics via \href{https://www.ons.gov.uk/methodology/geography/geographicalproducts/ruralurbanclassifications/2021ruralurbanclassification}{ons.gov.uk}.

The ITL1 region lookup used in the analysis, \textit{Ward to Local Authority District to CTY to RGN to CTRY (December 2021) Lookup in the UK}, is available from the Office for National Statistics Geoportal via \href{https://geoportal.statistics.gov.uk/datasets/7315ebba66154363806867bcfd52417a_0/explore}{geoportal.statistics.gov.uk}.

The Index of Multiple Deprivation (IMD) 2025 is available from the UK Government’s \textit{English indices of deprivation 2025: statistical release} via \href{https://www.gov.uk/government/statistics/english-indices-of-deprivation-2025/english-indices-of-deprivation-2025-statistical-release}{gov.uk}.

Code will be made publicly available via a repository upon acceptance.

\section{Funding and acknowledgements}
Duncan Cook was supported by an ESRC Doctoral Training Partnership award (ES/P000738/1). John Aston was supported by the David \& Claudia Harding Foundation. The authors thank Dennis Grube for helpful conversations and comments.

\section{Conflicts of interest}
The authors declare that they have no competing interests.

\section{Use of AI}

AI-based tools were used to assist with code development. All analytical decisions and interpretations are the authors' own.



\end{document}